\documentclass[%
 twocolumn, aip,
 sd,%
 amsmath,amssymb,
reprint,%
]{revtex4-1}

\usepackage{graphicx}
\usepackage{dcolumn}
\usepackage{bm}
\usepackage{siunitx}
\usepackage[version=4]{mhchem}
\usepackage{xcolor}

\begin{document}

\title{Influence of stoichiometry on interfacial conductance in \ce{LaAlO3}/\ce{SrTiO3} grown by \SI{90}{\degree} off-axis sputtering}

\author{Chunhai Yin}
\affiliation{Huygens-Kamerlingh Onnes Laboratory, Leiden University, P.O. Box 9504, 2300RA Leiden, The Netherlands}
\author{Dileep Krishnan}
\affiliation{EMAT, University of Antwerp, Groenenborgerlaan 171, 2020 Antwerp, Belgium}
\author{Nicolas Gauquelin}
\affiliation{EMAT, University of Antwerp, Groenenborgerlaan 171, 2020 Antwerp, Belgium}
\author{Jo Verbeeck}
\affiliation{EMAT, University of Antwerp, Groenenborgerlaan 171, 2020 Antwerp, Belgium}
\author{Jan Aarts}
\affiliation{Huygens-Kamerlingh Onnes Laboratory, Leiden University, P.O. Box 9504, 2300RA Leiden, The Netherlands}

\begin{abstract}
We report on the fabrication of conducting interfaces between \ce{LaAlO3} and \ce{SrTiO3} by \SI{90}{\degree} off-axis
sputtering in an Ar atmosphere. At a growth pressure of 0.04~mbar the interface is metallic, with a carrier density of
the order of $10^{13}$~cm$^{-2}$ at 3~K. By increasing the growth pressure, we observe an increase of the out-of-plane
lattice constants of the LaAlO$_3$ films while the in-plane lattice constants do not change. Also, the low-temperature
sheet resistance increases with increasing growth pressure, leading to an insulating interface when the growth pressure
reaches 0.10~mbar. We attribute the structural variations to an increase of the La/Al ratio, which also explains the
transition from metallic behavior to insulating behavior of the interfaces. Our research emphasizes the key role of the
cation stoichiometry of LaAlO$_3$ in the formation of the conducting interface, and also the control which is furnished
by the Ar pressure in the growth process.
\end{abstract}

\maketitle

The discovery of a high mobility conducting interface between \ce{LaAlO3} (LAO) and \ce{SrTiO3} (STO) has given rise to
numerous investigations \cite{Ohtomo2004}. This two-dimensional electron system (2DES) exhibits multiple intriguing
physical properties, such as superconductivity \cite{reyren2007}, magnetism \cite{li2011,bert2011,lee2013NM}, and gate
tunable insulator to metal \cite{thiel2006} and insulator to superconductor transitions \cite{caviglia2008N}. However,
the origin of the 2DES in still under debate. Proposed explanations basically fall into two classes, intrinsic charge
transfer and extrinsic defects mechanisms. The intrinsic mechanism considers the polar discontinuity between the polar
LAO and the nonpolar STO, which leads to a charge transfer above a critical thickness of LAO films \cite{nakagawa2006}.
The extrinsic mechanisms involve defects formed at the interface during the film deposition process, such as oxygen
vacancies in the STO substrate \cite{kalabukhov2007,siemons2007,herranz2007} and cation intermixing at the interface
\cite{willmott2007,chambers2010SSR}.

Pulsed laser deposition (PLD) is by far the most commonly used growth method to prepare LAO/STO interfaces. During the
PLD process,high energy particle bombardment could introduce the above defects into the interface, which makes it
difficult to understand the roles of the intrinsic and extrinsic mechanisms \cite{chambers2010SSR}. Other growth
techniques bring new insights here. \citet{Warusawithana2013} studied the LAO/STO system by molecular beam epitaxy
(MBE). The interesting outcome is that interfacial conductivity was only observed in Al-rich samples (La/Al $\leq$
0.97). Further density functional calculations demonstrated the different roles of defects in the charge transfer
mechanism. In Al-rich samples, Al can fill La vacancies without changing the net charge of the (001) planes. The
electronic reconstruction can still transfer electrons to the interface. In La-rich sample, however, La can not
substitute for Al, resulting in the formation of \ce{Al2O3}-vacancy complexes which prohibits the charge transfer.

Sputtering also has been used. High-pressure (\SI{1}{\milli\bar}) on-axis sputtering yielded LAO films with a La/Al
ratio of \SI{1.1}{},and insulating interfaces \cite{dildar2013}. \SI{90}{\degree} off-axis sputtering has been shown to
be capable of growing epitaxial and smooth films with conducting interfaces \cite{podkaminer2013}. Sputtering is widely
used in industry, which can also facilitate the device applications of LAO/STO interfaces. In this work, we show the
growth of high quality epitaxial LAO films by \SI{90}{\degree} off-axis sputtering. The La/Al ratio is tuned by varying
the growth pressure. As a consequence, we observe strong but controlled variations in the interfacial conductivity.

LAO films were grown on \ce{TiO2}-terminated STO (001) substrates. In order to obtain the \ce{TiO2} termination, the
substrates were etched by buffered \ce{HF} for 30 \si{\second} and annealed at \SI{980}{\degreeCelsius} in flowing
oxygen (\SI{150}{sccm}) for \SI{1}{\hour} \cite{koster1998}. In the sputtering system, the working distances were
\SI{75}{\milli\meter} from the surface of the heater to the axis of the target and \SI{45}{\milli\meter} from the
surface of the target to the axis of the heater. It should be noted that the proper choice of growth pressure is
strongly dependent on the working distances. A 2-inch single crystal LAO wafer was used as the sputtering target. The
growth temperature was \SI{800}{\degreeCelsius} and the RF power was \SI{50}{\watt}. Five samples were grown at various
Ar pressures from \SI{0.04}{\milli\bar} to \SI{0.10}{\milli\bar} (see Table \ref{table1}). In the following the samples
will be referred to with their growth pressure. The target was pre-sputtered for at least \SI{15}{\minute} in order to
stabilize an oxygen background partial pressure produced by the target \cite{podkaminer2013}. After deposition, the
samples were $ in\,situ $ annealed in \SI{1}{\milli\bar} oxygen at \SI{600}{\degreeCelsius} for \SI{1}{\hour} to remove
the oxygen vacancies in the STO substrates. The samples were then cooled down to room temperature in the same oxygen
atmosphere at a rate of \SI{10}{\degreeCelsius}/\si{\minute}. The deposition rate decreases from \SI{4.27}{\AA/\minute}
at \SI{0.04}{\milli\bar} to \SI{3.20}{\AA/\minute} at \SI{0.10}{\milli\bar}. Two reference samples were prepared to
test the effectiveness of the oxygen annealing treatment. One sample is a bare STO substrate heated up to the growth
temperature without film deposition. The other sample is an amorphous LAO/STO sample grown at room temperature at
\SI{0.08}{\milli\bar}. Both samples were highly conductive, which indicates the presence of oxygen vacancies
\cite{schooley1964PRL,chen2011nl}. The samples then underwent the above oxygen annealing treatment and were found to be
insulating.
\begin{figure}[t]
	\centering
	\includegraphics[width=0.9\linewidth]{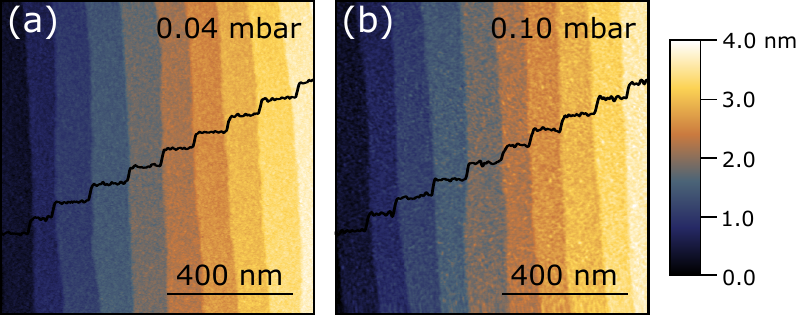}	
	\caption{AFM images of samples (a) 004 and (b) 010, using color code for the height. Insets are the height profiles of the surfaces.}
	\label{figafm}
\end{figure}

\begin{figure}[b]
	\centering
	\includegraphics[width=0.9\linewidth]{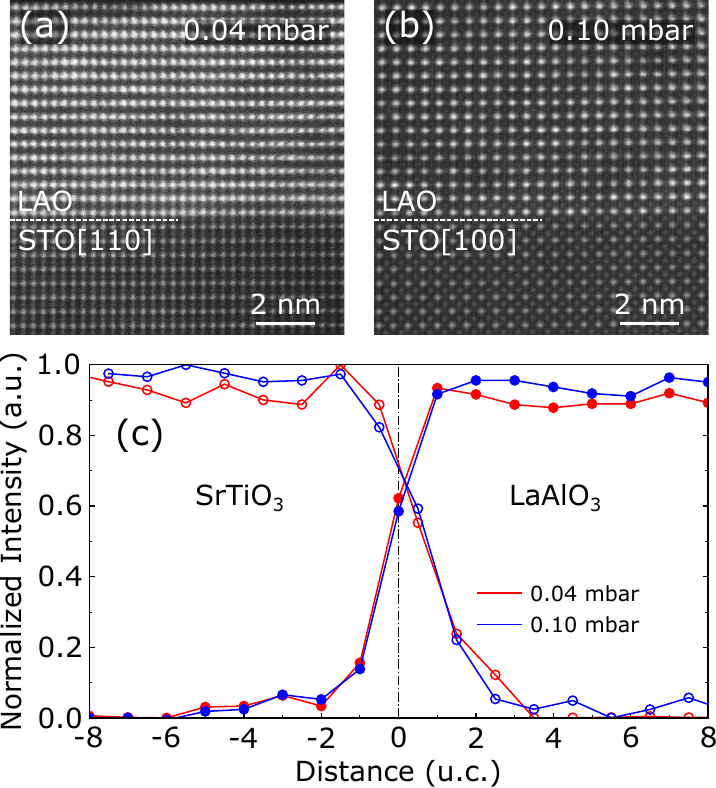}
	\caption{High-angle annular dark field STEM (HAADF-STEM) images of (a) sample 004 taken along the [110] direction and (b) sample 010
taken along the [100] direction. (c) STEM electron energy loss spectroscopy (STEM-EELS) analysis of samples 004 and 010, the La-M$_{4,5}$
(solid circles) and Ti-L$_{2,3}$ (open circles) edges integrated unit cell by unit cell across the interface.}
	\label{figtem}
\end{figure}
Surface topologies were measured by tapping mode atomic force microscopy (AFM). The epitaxial quality of the interface
was characterized by scanning transmission electron microscopy (STEM). Film thicknesses and lattice constants were
determined by high-resolution X-ray diffraction (HRXRD).  Magnetrotransport properties were measured with a Quantum
Design physical property measurement system (PPMS) by sweeping the magnetic field between $\pm$ \SI{9}{\tesla}. The
measurements were performed in the van der Pauw geometry. Ohmic contacts were formed by wedge bonding Al wire directly
to the sample surface.

Fig. \ref{figafm}(a) and Fig. \ref{figafm}(b) show the AFM topographic images of samples 004 and 010. An atomically
flat surface with clear step-and-terrace structure can be observed. The inset shows the step height which corresponds
to the STO (001) interplanar distance ($ \approx $ \SI{3.905}{\angstrom}). The epitaxial quality of the films was
further characterized by high-angle annular dark field STEM (HAADF-STEM). As shown in Fig. \ref{figtem}(a) and Fig.
\ref{figtem}(b), atomically sharp interfaces between the film and the substrate are clearly visible. Fig.
\ref{figtem}(c) shows the STEM electron energy loss spectroscopy (STEM-EELS) analysis of samples 004 and 010. This
concentration profile is obtained by integration of the EELS intensity of the La-M$_{4,5}$ and Ti-L$_{2,3}$ edges
during a spectrum image unit cell by unit cell in the growth direction. The profile is normalized by the maximum of
intensity and cation vacancies are neglected. Identical intermixing (4 unit cells) was observed for both samples. This
demonstrates that interdiffusion is a phenomenon that is not influenced by the growth pressure of the film.

\begin{figure}[t]
	\centering
	\includegraphics[width=0.9\linewidth]{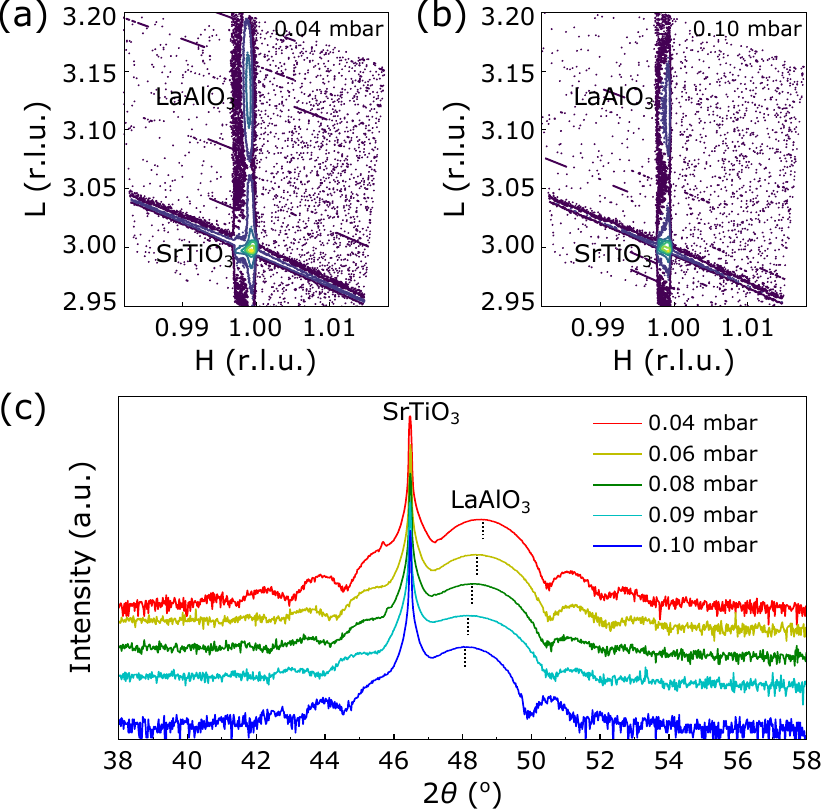}
	\caption{Reciprocal space maps (RSM) around the STO (103) diffraction peak of samples (a) 004 and (b) 010. (c) The $ \theta $-2$ \theta $
scans for samples grown at various Ar pressures. The dashed lines are the LAO (002) diffraction peaks.}
	\label{figxrd}
\end{figure}

Fig. \ref{figxrd}(a) and Fig. \ref{figxrd}(b) show the reciprocal space maps (RSM) around the STO (103) diffraction
peak of samples 004 and 010. The films are coherently strained to the substrate, which means that in-plane lattice
constant ($a_{\textrm{LAO}}$) is \SI{3.905}{\angstrom}. Fig. \ref{figxrd}(c) shows the $ \theta $-2$ \theta $ scans.
The dashed lines are the positions of LAO (002) diffraction peaks. It  can be seen that as the growth pressure
increases, the LAO peak shifts to lower angle, which corresponds to an increase of the out-of-plane lattice constant
($c_{\textrm{LAO}}$) \cite{qiao2011PRB}. By fitting the interference fringes, we extract $c_{\textrm{LAO}}$ as well as
the film thickness ($t_{\textrm{LAO}}$). Table \ref{table1} summarizes the estimated values for $a_{\textrm{LAO}}$,
$c_{\textrm{LAO}}$ and $t_{\textrm{LAO}}$ of the samples. It has been reported that the increase of $c_{\textrm{LAO}}$
is due to the increase of the La/Al ratio in LAO films. The relationship between them was systematically studied by
\citet{qiao2011PRB}. Thus, we extract the La/Al ratios of our samples by interpolating our data points using their
published results. The interpolated La/Al ratios are listed in Table \ref{table1}. As the growth pressure increases
from 0.04 mbar to 0.10 mbar, the La/Al ratio increases from 0.88 to 1.00.

\begin{table}[t]
	\centering
	\caption{Growth pressure, in-plane lattice constant ($a_{\textrm{LAO}}$), out-of-plane lattice constant ($c_{\textrm{LAO}}$),
thickness ($t_{\textrm{LAO}}$) and La/Al ratio of LAO films.}
	\setlength{\tabcolsep}{5pt}
	\renewcommand{\arraystretch}{1.2}
	\begin{tabular}{ccccc}
		\hline\hline
		Growth Pressure & $a_{\textrm{LAO}}$ & $c_{\textrm{LAO}}$ & $t_{\textrm{LAO}}$ & La/Al ratio\footnote{Interpolated
La/Al ratios from Ref. \cite{qiao2011PRB}.}\\
		(mbar) &  (\AA) &  (\AA) & (u.c.) & \\
ratio\footnote{Expected La/Al ratios from Ref. \cite{qiao2011PRB}.}\\
		\hline
		0.04 & 3.905 & 3.734 & 16 & 0.88     \\
		0.06 & 3.905 & 3.739 & 15 & 0.89     \\
		0.08 & 3.905 & 3.745 & 15 & 0.91     \\
		0.09 & 3.905 & 3.751 & 14 & 0.94     \\
		0.10 & 3.905 & 3.763 & 17 & 1.00     \\		
		\hline\hline
	\end{tabular}
	\label{table1}
\end{table}

\begin{figure}[t]
	\centering
	\includegraphics[width=0.9\linewidth]{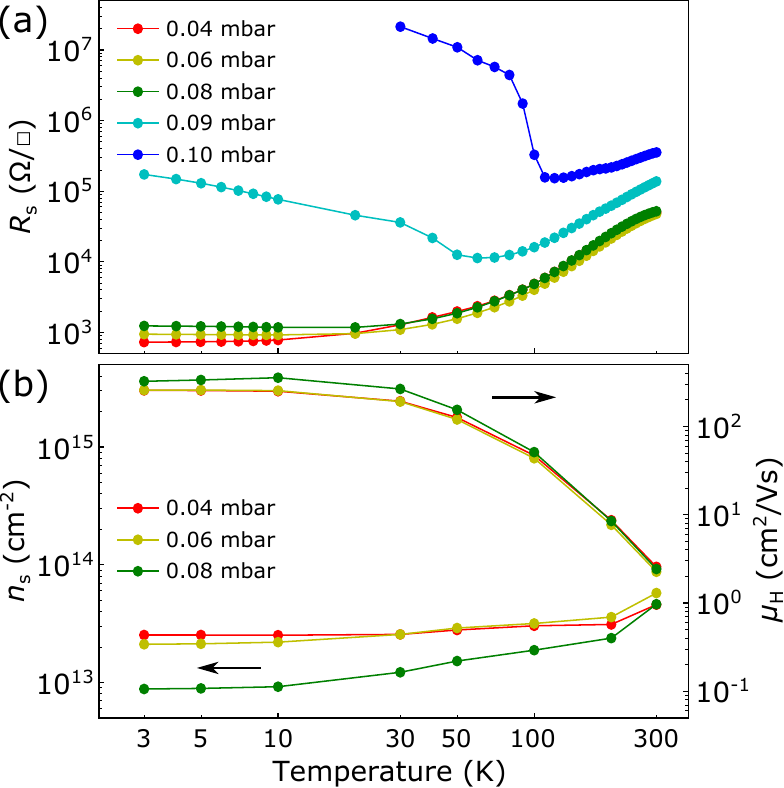}
	\caption{Temperature dependence of (a) the sheet resistance ($R_\textrm{s}$) and (b) the carrier density ($n_\textrm{s}$)
and the Hall mobility ($\mu_\textrm{H}$) for samples grown at various Ar pressures.}
	\label{figtrans}
\end{figure}

Fig. \ref{figtrans}(a) shows the temperature dependence of the sheet resistance ($R_\textrm{s}$) for samples grown at
various Ar pressures. Samples 004, 006 and 008 show similar metallic behavior from \SI{300}{\kelvin} to
\SI{3}{\kelvin}. The interfacial conductivity changes dramatically as the growth pressure further increases. For sample
009, $R_\textrm{s}$ decreases from \SI{1.4d5}{\ohm}/$\square$ at \SI{300}{\kelvin} to \SI{1.1d4}{\ohm}/$\square$ at
\SI{60}{\kelvin} and then gradually increases to \SI{2.2d5}{\ohm}/$\square$ at \SI{3}{\kelvin}. For sample 010,
$R_\textrm{s}$ decreases from \SI{3.5d5}{\ohm}/$\square$ at \SI{300}{\kelvin} to \SI{1.6d5}{\ohm}/$\square$ at
\SI{100}{\kelvin} and abruptly changes to insulating state afterwards. The temperature dependence of the carrier
density ($n_\textrm{s}$) and the Hall mobility ($\mu_\textrm{H}$) for the metallic samples are shown in Fig.
\ref{figtrans}(b). $n_\textrm{s}$ and $\mu_\textrm{H}$ were determined by \(n_\textrm{s}=1/eR_\textrm{H}\) and
\(\mu_\textrm{H}=R_\textrm{H}/R_\textrm{s}\), where $e$ and $R_\textrm{H}$ are the electron charge and the Hall
coefficient, respectively. $n_\textrm{s}$ and $\mu_\textrm{H}$ are approximately \SI{1d13}{\centi\meter^{-2}} and
\SI{2.6d2}{\centi\meter^{2}/\volt\second}, respectively, at \SI{3}{\kelvin}, which is consistent with reported results
of LAO/STO interfaces grown by sputtering \cite{podkaminer2013} and PLD \cite{kalabukhov2007,liu2013PRX}.

Our experimental results help to gain some insight in the role of extrinsic defects induced by the PLD process. First,
as is well known, oxygen vacancies can lead to conductivity in either bare STO substrates \cite{schooley1964PRL} or
LAO/STO interfaces \cite{cancellieri2010EPL,chen2011nl,liu2013PRX}. Our samples were grown in a reducing atmosphere,
thus there is a large amount of oxygen vacancies in STO without post oxygen annealing. The behavior of our reference
samples indicates that the post oxygen annealing treatment is efficient enough to remove the oxygen vacancies in the
STO substrate created during the film deposition.

Second, it has been reported that La-doped STO shows metallic behavior \cite{marina2002SSI}. At the LAO/STO interface,
La/Sr intermixing could be induced in two ways. One way is simply by the PLD process itself, during which the STO
substrate is bombarded by particles with kinetic energies around several tens of \si{\electronvolt}
\cite{chambers2010SSR}. In our off-axis sputtering deposition, we use relatively high Ar pressures (0.04-0.10
\si{\milli\bar}), which correspond to mean free paths of several millimeters. The direct distance between the center of
the target and the substrate is about \SI{87.5}{\milli\meter}. The ejected particles would undergo multiple scatterings
to slow down their speed before they deposit on the substrate. In our case, the chance of introducing La/Sr intermixing
by high energy particle bombarding should be low. The other way is the dipole compensation mechanism proposed by
\citet{nakagawa2006}, where a compensating dipole is produced by La/Sr intermixing to reduce the interface dipole
energy. We observed identical intermixing in samples 004 and 010. However, intermixing should not be the origin of
conductivity otherwise the two samples would show similar conducting behavior. We therefore conclude that oxygen
vacancies or cation intermixing may exist at our LAO/STO interfaces, however, their contributions to the conductivity
are negligible.

In our experiments, all the LAO films are epitaxially strained to STO substrates. Increasing the growth pressure only
increases the La/Al ratio, which we believe is due to light Al being scattered more easily at higher pressures
\cite{dildar2013}. The dramatic change in the transport properties is related to the change of cation stoichiometry of
the LAO films. Thus our results agree with the cation stoichiometry mechanism proposed by \citet{Warusawithana2013}.
For the LAO/STO samples grown by PLD, it has been reported that a slight variation in growth parameters modifies the
cation stoichiometry of LAO \cite{golalikhani2013JAP,breckenfeld2013PRL,sato2013APL}, also resulting in a dramatic
change in the interfacial conductivity. However, the cation stoichiometry is not checked on a routine basis. It might
explain the fact that samples from different PLD groups are often hardly comparable, although similar growth parameters
are used.

In conclusion, high quality epitaxial LAO films were grown on STO (001) substrates by \SI{90}{\degree} off-axis
sputtering. While increasing the growth pressure, little structural variations have been observed, except for an
increase of the out-of-plane lattice constant, which indicates an increase of the La/Al ratio. Metallic conducting
interfaces were only found in Al-rich samples. Our results emphasize that cation stoichiometry in LAO films plays an
important role in the formation of interfacial conductivity at the LAO/STO interfaces.

We thank Nikita Lebedev, Aymen Ben Hamida and Prateek Kumar for useful discussions and Giordano Mattoni, Jun Wang,
Vincent Joly and Hozanna Miro for their technical assistance. We also thank Jean-Marc Triscone and his group for
sharing their design of the sputtering system with us. This work is supported by the Netherlands Organisation for
Scientific Research (NWO). C. Yin is supported by China Scholarship Council (CSC) with grant No. 201508110214. N.G.,
D.K. and J.V. acknowledge financial support from the GOA project ``Solarpaint'' of the University of Antwerp.

\bibliographystyle{apsrev4-1}
\bibliography{LAO1}
\end{document}